\pdfoutput=1
\documentclass[preprint,aps]{revtex4}

\usepackage{graphicx}
\usepackage{dcolumn}
\usepackage{bm}

\input{tcilatex}
\begin{document}

\title{Coulomb implosion mechanism of negative ion acceleration in laser
plasmas}
\author{T. Nakamura$^{\ast }$, Y. Fukuda, A. Yogo, M. Tampo,\ M. Kando, Y.
Hayashi, T. Kameshima, A. S. Pirozhkov, T. Zh. Esirkepov, T. A. Pikuz, A.
Ya. Faenov, H. Daido, and S. V. Bulanov}
\affiliation{}
\date{\today }

\begin{abstract}
Coulomb implosion mechanism of the negatively charged ion acceleration in
laser plasmas is proposed. When a cluster target is irradiated by an intense
laser pulse and the Coulomb explosion of positively charged ions occurs, the
negative ions are accelerated inward. The maximum energy of negative ions is
several times lower than that of positive ions. The theoretical
description and Particle-in-Cell simulation of the Coulomb implosion
mechanism and the evidence of the negative ion acceleration in the
experiments on the high intensity laser pulse interaction with the cluster
targets are presented.
\end{abstract}

\maketitle

Recent developments in the ultra-intense laser pulse technology resulted in
the intensity at the level above 10$^{22}$ W/cm$^{2}$ \cite{Yan}. The ultra-intense laser pulse interaction with matter opens such new research fields as the fast
ignition of inertial thermonuclear fusion \cite{FI}, the charged particle
beam acceleration for medical applications \cite{medical}, and the
development of compact sources of high energy electrons, ions and photons 
\cite{ele,ion,xray} (see also the review article \cite{MTB} and literature
quoted in).

The negative ion generation has attracted a great deal of  attention due to
various applications as the beam injectors for magnetically confined fusion
devices \cite{MCF}, as the heavy ion fusion drivers \cite{HIB}, and for
utilizing them in large scale ion accelerators \cite{JPARC}. Negatively
charged ion beams are typically generated via the charge exchange process
during the positive ion propagation through the alkali metal vapor. However,
since the charge exchange cross section rapidly decreases as the incident
energy exceeds keV level, it has been an important issue to find an
effective way to generate the high energy negative ions. Laser plasma 
provides a source of the negative ions with the energy in keV$\sim$MeV
range \cite{Sargis,Volkov,Yogo}. In the experiment presented in Ref. \cite%
{Sargis}, the negative ions with the energy about  MeV have been observed
when the water droplets are irradiated by the ultra-short and ultra-intense
laser pulse (40 fs, $I\approx 10^{19}$ W/cm$^{2}$). The negative ions were generated
inside the laser plasma and are accelerated together with the positive ions.
As far as it concerns the positive ion acceleration in the laser plasmas,
there were invoked several mechanisms including the sheath acceleration 
\cite{TNSA}, the Coulomb explosion \cite{Coulomb}, the shock acceleration 
\cite{shock}, the radiation pressure dominant acceleration \cite{piston},
the ``after-burner" \cite{BOA}, etc. On the other hand, the acceleration
mechanism of negative ions, according to our knowledge, has not yet been
studied so far.

In the present Letter we propose the Coulomb implosion mechanism of the
negative ion acceleration. When a cluster target is irradiated by an intense
laser pulse and the\ ponderomotive pressure of the laser light blows away
the electrons, the repelling force of an uncompensated electric charge of
positive ions causes the cluster Coulomb explosion. In the case of a multi-species
cluster with a relatively small number of the negative ions, where possible mechanisms of negative ion creation are discussed 
in Ref.\cite{Smirnov}, the electric
field formed by the positively charged component accelerates the negative
ions inward. The negative ions leave the target, passing through the center or bouncing its vicinity. Below we formulate the theoretical description and present
the results of the Particle-in-Cell simulation of the Coulomb
explosion/implosion of a multi-species cluster. The evidence of the negative
ion acceleration in the experiments on the high intensity laser pulse
interaction with the cluster targets is presented as well.

At first, we consider the Coulomb explosion dynamics of a multi-species
cluster. The cluster comprises the positive and negative ion components.
Within the framework of hydrodynamics description  the positive and negative
ion motion is governed by the continuity equations: $\partial _{t}n_{\pm
}+\nabla \cdot (n_{\pm }\mathbf{v}_{\pm })=0$, the Euler equations: $%
\partial _{t}\mathbf{v}_{\pm }+\left( \mathbf{v}_{\pm }\nabla \right) 
\mathbf{v}_{\pm }=\pm e_{\pm }\mathbf{E/}M_{\pm }$, and by the equation for the
electric field: $\nabla \cdot \mathbf{E}=4\pi \sum(\pm e_{\pm
}n_{\pm })$. Here $\mathbf{E}$ is the electric field, the subscript $(\pm )$
stands for positive and negative ions, $n_{\pm }$, $\mathbf{v}_{\pm }$, $%
\pm e_{\pm }$, and $M_{\pm }$ are their density, velocity, electric charge and
mass. These equations admit a self-similar solution with homogeneous
deformation, for which the densities are homogeneous, $n_{\pm }=n_{\pm }(t)$,
 and the velocities and electric field are linear functions of the
coordinates: $v_{i,\pm }=\dot{m}_{ik,\pm }(t)\left( m_{kj,\pm }\right)
^{-1}(t)x_{j}$, $E_{i}=\epsilon _{ij}(t)x_{j}$. Here summation over repeated
indices is assumed. The matrix $m_{ik,\pm }$ is the deformation matrix, $%
\dot{m}_{ik,\pm }$ is its time derivative and $\left( m_{kj,\pm }\right)
^{-1}$ is its inverse matrix (see Ref. \cite{Sedov}). Now we assume a spherical, cylindrical or
planar symmetry of the flow, for which the deformation matrix has a diagonal
form with equal diagonal elements, $k_{\pm }(t)$. In this case, the velocity of 
Lagrange element is given by $v_{\pm}=\dot{k_{\pm}}x^0$ and of the 
Euler element by $v_{\pm}=\dot{k_{\pm}}x/k_{\pm}$, and the 
density is $n_{\pm}=n_{0,{\pm}}/k_{\pm}^d$, with $n_{0,+}$ and $n_{0,-}$ being the initial
values of densities of the positive and negative ions; $d$ equals to the
dimension such that $d=1$, $d=2$ and $d=3$ correspond to planar,
cylindrical, and spherical geometry, respectively. Substituting these
functions to the above written equations, we obtain for $k_{\pm }(t)$ a
system of two nonlinear ordinary differential equations:%
\begin{eqnarray}
\frac{d^{2}k_{+}}{dt^{2}}&=&\frac{\omega _{+}^{2}}{k_{+}^{(d-1)}}-\frac{\omega
_{\pm }^{2}k_{+}}{k_{-}^{d}},  \label{1}\\
\frac{d^{2}k_{-}}{dt^{2}}&=&\frac{\omega _{-}^{2}}{k_{-}^{(d-1)}}-\frac{\omega
_{\mp }^{2}k_{-}}{k_{+}^{d}},  \label{2}
\end{eqnarray}%
where we introduced $\omega _{+}=\sqrt{4\pi n_{0,+}e_{+}^{2}/M_{+}}$, $%
\omega _{-}=\sqrt{4\pi n_{0,-}e_{-}^{2}/M_{-}}$, $\omega _{\pm }=\sqrt{4\pi
n_{0,-}e_{+}e_{-}/M_{+}}$, and $\omega _{\mp }=\sqrt{4\pi
n_{0,+}e_{+}e_{-}/M_{-}}$.

If the density of negative ions is small enough, $n_{0,-}\ll
e_{+}n_{0,+}/e_{-}$, then we assume it vanishes, \ $n_{0,-}=0$, at the first
step of approximation with respect to a small parameter $%
e_{-}n_{0,-}/e_{+}n_{0,+}$. In this case for $d=3$, Eq. (\ref{1}) describes
the Coulomb explosion of a positively charged cluster \cite{L-J}. For
initial conditions, $k_{+}(0)=1$ and $dk_{+}/dt(0)=0$, its solution reads $%
\sqrt{k_{+}(k_{+}-1)}+\log \left( \sqrt{k_{+}}+\sqrt{k_{+}-1}\right) =\sqrt{2%
}\omega _{+}t$. \ In the case of cylindrical geometry, $d=2$, which may
correspond to the Coulomb explosion of the self-focusing channel \cite{SAR},
Eq. (\ref{1}) yields $\mathrm{Erfi}\left( \sqrt{\log k_{+}}\right) =\sqrt{%
2/\pi }\omega _{+}t$. Here $\mathrm{Erfi}\left( z\right) =\mathrm{erf%
}(iz)/i$ is the complex error function which is a real function of its
argument. In a planar geometry, $d=1$, we have $k_{+}=1+(\omega _{+}t)^{2}/2$.

Then, assuming the dependence of $k_{+}$ on time to be given by the above
written expressions, we obtain from Eq. (\ref{2}) the equation $%
d^{2}k_{-}/dt^{2}+\Omega _{-}^{2}(t)k_{-}=\omega _{-}^{2}/k_{-}^{(d-1)}$
with $\Omega _{-}(t)=\omega _{\mp }/(k_{+}(t))^{d/2}$ and with initial
conditions $k_{-}(0)=1$ and $dk_{-}/dt(0)=0$. If $\omega _{-}/\Omega _{-}\ll
1$, the negative ions during the first step move towards the center from
where they bounce and move outwards. The ion velocity can be estimated to
depend on the initial position, $x^{0}$, as $|v_{-}|=\left(
\tint\nolimits_{0}^{\infty }\Omega _{-}(t^{\prime })dt^{\prime }\right) x^{0}
$. In the opposite limit, when $\omega _{-}/\Omega _{-}\gg 1$, we seek a
solution in the form of slowly varying and fast oscillating components, $%
k_{-}(t)=K_{-}(t)+\tilde{k}_{-}(t)$, with $\left\langle \tilde{k}%
_{-}\right\rangle =0$, where $<...>$ denotes a time averaging. The
component $K_{-}$ corresponds to the equilibrium solution of Eq. (\ref{2})
for frozen dependence on time of $k_{+}$, i.e. $K_{-}(t)=\left( \omega
_{-}/\omega _{\mp }\right) ^{2/d}k_{+}(t)$. For oscillating component, $%
\tilde{k}_{-}$, we find the equation of an oscillator with a time dependent
frequency: $d^{2}k_{-}/dt^{2}+d\Omega _{-}^{2}(t)k_{-}=0$. Due to a
smallness of the ratio $\omega _{-}/\omega _{\mp }\ll 1$ a dependence on
time of $\Omega _{-}(t)$ is slow and we can use the WKB approximation. This
yields the expression for the oscillating part of $k_{-}(t)$. It reads $%
\tilde{k}_{-}(t)=\left[ k_{+}(t)\right] ^{d/4}\cos \left( \sqrt{d}%
\int_{0}^{t}\Omega _{-}(t^{\prime })dt^{\prime }\right) $, which describes
the particle oscillations with growing amplitude and decreasing frequency.

The ion energy spectrum can be found by considering a dependence of the
velocity on the ion initial position. We see that the ion kinetic energy, $%
\mathcal{E}=M_{-}v_{-}^{2}/2$, is proportional to the square of the ion
initial coordinate: $\mathcal{E}\propto (x^{0})^{2}$. Using a condition of
the continuity in the phase space, $dN_{-}=4\pi n_{0,-}(x^{0})^{d-1}d%
\mathcal{E}$/$|d\mathcal{E}/dx^{0}|$ we find that the ion energy spectrum is
inversely proportional to the square root of energy, $dN/d\mathcal{E}\propto 1/%
\sqrt{\mathcal{E}}$ in the case of planar geometry; it is flat, $dN/d%
\mathcal{E}=\mathrm{constant}$ in cylindrical geometry, and is proportional
to the square root of energy $\propto \sqrt{\mathcal{E}}$ in the case of
spherical geometry. The negative ion energy is in a factor $\kappa =\left(
\omega _{-}/\omega _{\mp }\right) ^{4/d}=(e_{-}n_{0,-}/e_{+}n_{0,+})^{2/d}$
smaller than the energy of the positive ions.

Now we demonstrate the implosion dynamics of negative ions by using the
two-dimensional Particle-in-Cell (PIC) simulations. The simulation
conditions are as follows. The target is a cluster with the diameter of 1 $%
\mu $m  composed of electrons, protons and negative hydrogen ($H^{-}$) ions
whose densities are 1.0$n_{c}$, 1.1$n_{c}$, and 0.1$n_{c}$, respectively.
Here $n_{c}=m_{e}\omega ^{2}/4\pi e^{2}$ is the critical density for the
laser light with the frequency $\omega $. In this case the parameter $\kappa 
$ equals 0.1. The electron initial temperature is set to be equal to 500 eV.
The ions are assumed to be initially cold. The target is located at the
center of the simulation box which has a size of 40$\mu $m in both the $x-$
and $y-$directions. The laser pulse irradiates the target from the left hand
side boundary. It propagates in the $x-$ direction and is polarized along
the $y-$ direction. The pulse has a flat top form rising up in a laser
period and keeping its peak intensity of $1.0\times 10^{20}$ W/cm$^{2}$ which is constant for the duration of 15 fs, where laser wavelength is 1 $\mu$m. The laser pulse is focused to the spot
with the diameter of 3$\mu $m (FWHM), where the focal point is $x=18 \mu$m. The simulation
box has 4000$\times $4000 meshes, with $\simeq 1\times 10^{6}$
quasi-particles used.

\begin{figure}[tbp]
\includegraphics[width=8.5cm]{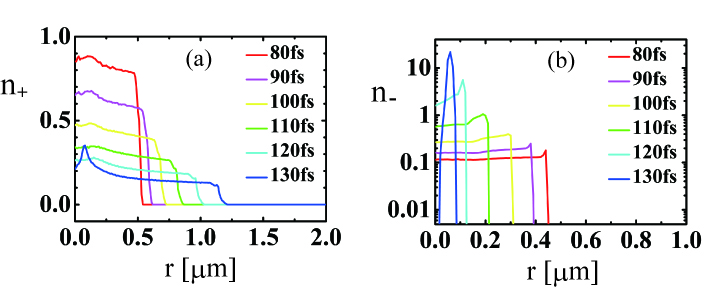}
\caption{ Radial density profiles of (a) protons and (b) negative hydrogen, H$^{-}$, ions during
the Coulomb explosion and implosion. The densities are normalized on the
critical density.}
\label{fig:1}
\vspace{-0.2cm}
\end{figure}
\begin{figure}[tbp]
\includegraphics[width=8.5cm]{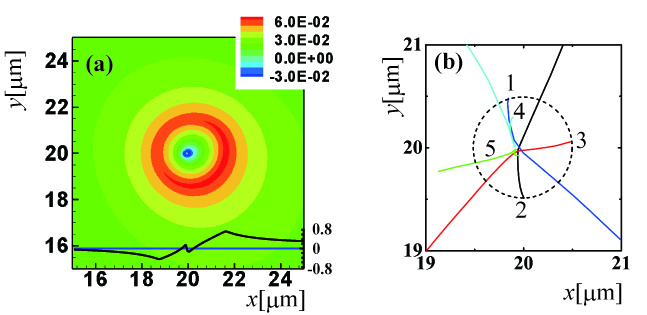}
\caption{(a) Distribution of the radial component of electric field $E_r$ at
time t=130 fs. The electric field is normalized on the incident laser electric field,
 which is 27TV/m in our simulations. The profile of $E_r$ along the horizontal line passing y=20 
 is shown in the lower part. (b) 
 Trajectories of $H^-$ ions.}
\label{fig:2}
\vspace{-0.3cm}
\end{figure}

By the irradiation of the ultra-intense laser pulse, the electrons are
heated and swept away from the target leaving the target positively charged.
This results in the Coulomb explosion of the positive ions and in the
Coulomb implosion for the negative ions. The time history of radial density
profile of protons and $H^{-}$ ions along x-axis (x$\leq $20) is plotted in
Fig. \ref{fig:1} (a) and (b), respectively ($r=-$x+20). The protons expand radially due
to the Coulomb explosion, having a rather flat distribution as in the
self-similar solution of two-dimensional case. $H^{-}$ ions start to move
inward from the outer region, and they are most compressed at $t=130$ fs
whose center is located at (x,y)=(19.95,20.0), which is 0.05 $\mu $m left to
the target center. This off-center of the implosion comes from the asymmetry
of laser irradiation. As the $H^{-}$ ions are accelerated towards the
center, the electric field, which sign is opposite to that of the outer
region, is induced at the center due to accumulation of the negative charges,
as it is seen in Fig.\ref{fig:2} (a). The electric field decelerates the
imploding $H^{-}$ ions and attract protons towards the center which is seen
in Fig.\ref{fig:1}(a). The trajectories of $H^{-}$ ions are shown in
Fig.\ref{fig:2}(b). Particles with numbers from 1 to 3 are initially located
at the periphery of the target. They are accelerated towards the center and
slightly deflected by the electric field and leave the target.
They have maximum energy of 1 MeV when they reach the cluster
center, and then they are gradually decelerated while leaving the target.
Particle number 4 moves towards the center and then is reflected back by the
induced strong electric field. Its energy at the time when it has left the
target is 0.1 MeV. Particle number 5 being located near the center bounces between the outer and inner regions and then leaves
the target. It achieves the energy being only $\sim$30 eV. As a result, energetic $H^{-}$ ions are generated from the periphery of the target. 
\begin{figure}[tbp]
\includegraphics[width=8.5cm]{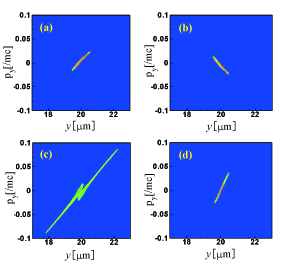}
\caption{Phase planes $(y,p_y)$ of the positive (in the frames (a) and (c)) and 
negative (in the frames (b) and (d)) ions.
The rows (a,b) and (c,d) correspond to the time t=80fs and t=180fs, respectively.}
\label{fig:3}
\end{figure}

The Coulomb explosion/implosion of positive/negative ions is illustrated in Fig. \ref{fig:3},
where the phase planes $(y,p_y)$ are plotted. In the row (a,b) we show the ion phase planes at 
time t=80 fs. We see the linear (self-similar) dependence of the ion momentum on the 
coordinate corresponding to the positive ion expansion and to the negative ion collapse.
In the row (c,d) we plot the phase planes at the time t=180 fs, after the negative ions have
bounced from the central region. We see that both the positive and negative ions expand radially.
A local compression of the positive ions induced by the electric field formed at the center is also seen in Fig. \ref{fig:3}c.

\begin{figure}[tbp]
\includegraphics[width=8.5cm]{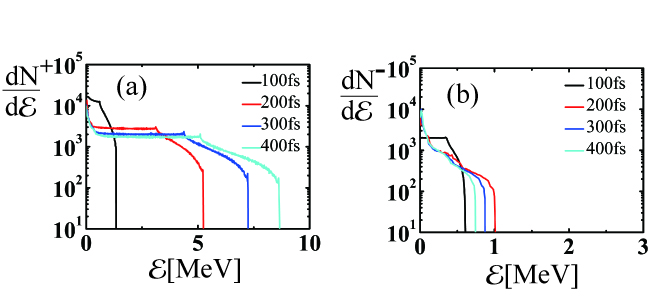}
\caption{Time history of the energy spectra of (a) protons, and (b) $H^{-}$
ions.}
\label{fig:4}
\vspace{-0.2cm}
\end{figure}

The energy spectra of protons and $H^{-}$ ions are shown
in Figs.\ref{fig:4}(a) and (b), respectively. The spectra of protons are
flat, which is in agreement with the theoretical prediction in the case of
two-dimensional Coulomb explosion. The maximum energy of negative hydrogen 
$H^{-}$ is equal to 1.0 MeV. We see that the negative ion energy is
approximately ten times less than the energy of positive ions. This is in
agreement with the above formulated theoretical model with 
$\mathcal{E}_{-}=\kappa \mathcal{E}_{+}$, for the parameter $\kappa $ equal to $0.1$ in
our simulations. 

In the case of lower values of the laser intensity or/and higher values of  total
initial number of the ions, the laser light does not blow away all the
electrons. For example, when a laser intensity is $I=1.0\times 10^{19}$W/cm$^2$ and a target has the same parameters as above calculations, we do not have a regime of pure Coulomb explosion.
Nevertheless, the negative ions are accelerated towards the target center.
Since the sheath field inside the expansion front is weakened by the bulk
electrons, the energy achieved by negative ions does not oscillate and they
sustain the energy after leaving the target.

\begin{figure}[tbp]
\includegraphics[width=8.5cm]{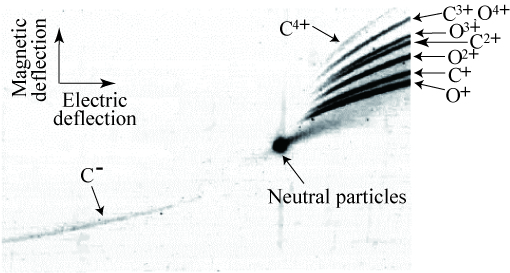}
\caption{A pattern registered on CR-39 with the Thomson parabola. }
\label{fig:5}
\end{figure}

Negative ions are observed in experiments using water droplet\cite{Sargis}, 
or thin foil target\cite{Volkov}. In our experiments negative ions are observed for the first time using mixtures of CO$_{2}$ clusters and He gas target. The laser has energy of 130 mJ with pulse duration of 35 fs. The pulse is focused into
the spot with 30 $\mu $m in diameter, resulting in the intensity of $7\times
10^{17}$ W/cm$^{2}$ in vacuum. The CO$_{2}$ clusters with the diameters equal to 0.35 $%
\mu $m, which contain $5\times 10^{8}$ molecules each, are generated using 
specially designed supersonic gas jet nozzle\cite{Fukuda}. Average distance between
the clusters is 5 $\mu $m. The laser pulse contrast of 10$^{-6}$ at the ns
time scale is expected to lead to the cluster heating and evaporation by the
pre-pulse. This results in the interaction of the main, fs, laser pulse with
the plasma clouds of larger volume and lower density than in the initial
clusters. The registered parabolic line of the positive and negative 
ions accelerated in the Coulomb explosion/implosion are shown in 
Fig.\ref{fig:5}. The Thomson parabola is positioned in the direction 
of 135$^{\circ}$ from the laser axis, i.e. we observe the ions moving in the backward-aside
direction with respect to the laser beam propagation. In Fig.\ref{fig:5} we
see the parabolic lines produced by negative C$^{-}$
ions as well as the carbon ions up to C$^{4+}$ ion
and by the oxygen ions up to O$^{4+}$ ions. The maximum energy of positive C$^{4+}$ ions is 4.8 MeV, and that of the negative C$^{-}$ ions is 0.6 MeV. The ratio of maximum energy of positive and 
negative ions is 1/8, which corresponds to $\kappa=1/8$ with $n_{0,-}/n_{0,+}\simeq 0.02$. 
The detailed discussion of experimental
results will be presented in the forthcoming papers \cite{Yogo}.

In conclusion, the proposed Coulomb implosion model explains the
acceleration of negatively charged ions in the laser-plasma. The negative
ions initially located at the target periphery are accelerated more
efficiently. A final energy of the negative ions is several times less than
the positive ion energy. The Coulomb implosion
mechanism is clearly demonstrated  by the PIC simulations. 
Acceleration of positive ions via the Coulomb explosion is a well-known mechanism. 
Negative ions can be accelerated in the same field, however, their acceleration occurs 
in the opposite direction with the bouncing-back in the vicinity of the center 
of symmetry, which results in the generation of high energy negative ions. 
We note here that the mechanism is applicable in the case of thin foil, self-focusing channel, and cluster according to in 1D, 2D, and 3D configuration.

We acknowledge a partial support from special Coordination Funds for
Promoting Science and Technology commissioned by MEXT of Japan. The
simulations are performed by using Altix3700 at JAEA Tokai.

$^{\ast }$ E-mail address : nakamura.tatsufumi@jaea.go.jp

\end{document}